\documentclass[usenatbib]{mnras}

\usepackage{mathptmx}
\usepackage{txfonts}

\usepackage[T1]{fontenc}
\usepackage{ae,aecompl}

\usepackage{graphicx}	
\usepackage{amssymb}	
\usepackage{gensymb}

\newcommand{\Msol}{\ensuremath{M_{\odot}\,}}
\newcommand{\Mbh}{\ensuremath{M_{BH}}}

\title[$\gamma$-ray emission characteristics of 3C~454.3]{Localising the $\gamma$-ray emission region during the June 2014 outburst of 3C~454.3}

\author[Rosemary T. Coogan, Anthony M. Brown and Paula M. Chadwick]{Rosemary T. Coogan$^{1}$\thanks{E-mail: r.coogan@sussex.ac.uk; now at the University of Sussex}, Anthony M.
Brown$^{1}$ and Paula M. Chadwick$^{1}$\\
$^{1}$ Durham University, Department of Physics, South Road, Durham, DH1 3LE, U.K}

\date{Accepted 2016 January 21. Received 2016 January 20; in original form 2015 June 30}

\pubyear{2016}

\begin{document}
\fontsize{9.2}{9.2}
\selectfont
\label{firstpage}
\pagerange{\pageref{firstpage}--\pageref{lastpage}}
\maketitle

\begin{abstract}
In May - July 2014, the flat spectrum radio quasar 3C~454.3 exhibited strong flaring behaviour. Observations with the Large Area Telescope detector on-board the \textit{Fermi} Gamma-ray Space Telescope captured the $\gamma$-ray flux at energies 0.1~$\leq E_{\gamma}\leq$~300~GeV increasing fivefold during this period, with two distinct peaks in emission.

The $\gamma$-ray emission is analysed in detail, in order to study the emission characteristics and put constraints on the location of the emission region. We explore variability in the spectral shape of 3C~454.3, search for evidence of a spectral cutoff, quantify the significance of VHE emission and investigate whether or not an energy-dependence of the emitting electron cooling exists. $\gamma$-ray intrinsic doubling timescales as small as $\tau_{int} = 0.68$ $\pm$ 0.01~h at a significance of >~5$\sigma$ are found, providing evidence of a compact emission region. Significant $E_{\gamma, emitted}\geq$~35~GeV and $E_{\gamma, emitted}\geq$ 50~GeV emission is also observed. The location of the emission region can be constrained to $r\geq1.3$~$\times$~$R_{BLR}^{out}$, a location outside the broad-line region. The spectral variation of 3C~454.3 also suggests that these flares may be originating further downstream of the supermassive black hole than the emission before and after the flares.
\end{abstract}

\begin{keywords}
gamma-rays: galaxies -- galaxies: active -- quasars: individual: 3C~454.3 -- galaxies: jets -- galaxies: nuclei
\end{keywords}



\section{Introduction}

Thanks to the launch of the \textit{Fermi} Gamma-ray Space Telescope in June 2008, we now have access to $\sim$7 years of $\gamma$-ray data from both Galactic and extragalactic sources. More than half of the $\gamma$-ray sources detected by the Large Area Telescope (LAT) on-board the \textit{Fermi} satellite are active galactic nuclei (AGN), with $\sim$98\% of these AGN a subclass known as blazars \citep{ref:Ackermann2015}. Blazars are oriented closely towards our line of sight, and can be further split into two subclasses - BL Lacertae (BL Lac) objects and flat spectrum radio quasars (FSRQs) \citep{ref:Urry1995}. The effect of this direct orientation is that the emission we see from blazars is highly Doppler boosted, causing blazars to appear as some of the brightest objects in the $\gamma$-ray sky, particularly during flaring episodes \citep{ref:Abdo2011}.

Despite the volume of $\gamma$-ray data that we are able to collect from both the \textit{Fermi}-LAT and ground-based instruments, we are unable to resolve the emission spatially. The process of locating the emission region is therefore indirect, and many different methods have previously been employed. The origin of the $\gamma$-ray emission from blazars has traditionally been assumed to be close to the central supermassive black hole (SMBH). This conclusion is based in part on the results of spectral energy distribution (SED) modelling (e.g. \citealt{ref:GeneralBlazars, ref:Nalewajko2012}), as well as the compact size of the emission region inferred from observations of rapid $\gamma$-ray variability \citep{ref:Tavecchio2010}. Using the size of the emission region to infer its location rests on the assumption of a constant jet geometry, where the size of the emission region, $R$, is related to the distance from the SMBH, $r$, and constant opening angle, $\Psi$, by $r\sim R/\Psi$ \citep{ref:Dermer2009, ref:GhiselliniTavecchio2009}. There are, however, studies that have concluded a molecular torus (MT) or parsec-scale origin for the $\gamma$-ray emission from blazars \citep{ref:Lahteenmaki2003, ref:Marscher2010, ref:Agudo2011, ref:Jorstad2010, ref:Jorstad2013}. Multi-wavelength (MWL) studies of blazars have resolved outbursts in radio emission on a parsec-scale from the SMBH, and simultaneous flares in the $\gamma$-ray regime suggest a common origin for the $\gamma$-ray emission \citep{ref:Marscher2010}. The presence of significant very high energy (VHE) emission from blazars also supports the idea that the emission region is not located within the broad-line region (BLR) \citep{ref:Donea2003, ref:Lui2006}. The possibility of multiple emission regions has recently been suggested, based on $\gamma$-ray observations (e.g. \citealt{ref:Brown1510}).

The FSRQ 3C~454.3 has been extraordinarily bright over the past decade. In December 2009, 3C~454.3 reached a record high energy $\gamma$-ray flux level for blazars, with a daily flux ($E>100$~MeV) of $F~=~(2.2~\pm~0.1)~\times~10^{-5}$~ph~cm$^{-2}$~s$^{-1}$ \citep{ref:Ackermann2010} and $F~=~(2.0~\pm~0.4)~\times~10^{-5}$~ph~cm$^{-2}$~s$^{-1}$ \citep{ref:Striani2010} observed by \textit{Fermi} and \textit{AGILE} \citep{ref:Tavani2009} respectively. It flared spectacularly again in November 2010, becoming brighter than even the Galactic Vela pulsar. The daily flux measured for this flare peaked at $F~=~(6.6~\pm~0.2)~\times10^{-5}$~ph~cm$^{-2}$~s$^{-1}$ \citep{ref:Abdo2011}, with a flux of $F~=~(6.8~\pm~1.0)~\times10^{-5}$~ph~cm$^{-2}$~s$^{-1}$ detected on a timescale of $\sim$12~h \citep{ref:Vercellone2011}. The analysis in 3 hour time bins revealed that the flux reached $F=(8.5\pm0.5)~\times~10^{-5}$~ph~cm$^{-2}$~s$^{-1}$ on Modified Julian Date (MJD) 55520 \citep{ref:Abdo2011}. These high flux levels have enabled extensive analysis to be done on the $\gamma$-ray characteristics of 3C~454.3, and the $\gamma$-ray emission has been suggested to originate both from the BLR and on parsec-scale distances from the SMBH (e.g. \citealt{ref:Ackermann2010, ref:Tavecchio2010, ref:Vercellone2010, ref:Vercellone2011, ref:Abdo2011, ref:Bonnoli2011, ref:Jorstad2013, ref:Vittorini2014}). 3C~454.3 has also been seen to flare brightly in the optical and radio \citep{ref:Villata2007, ref:Raiteri2008, ref:Hagen-Thorn2009, ref:Jorstad2010, ref:Vercellone2011}.

In this paper, we study in detail the $\gamma$-ray flares peaking in June 2014 from 3C~454.3, in order to understand more deeply the characteristics and location of the $\gamma$-ray emission. We assume a leptonic origin from a spherical emission region, where high energy electrons in the relativistic jet up-scatter low energy photons external to the jet, through inverse-Compton (IC) scattering \citep{ref:GeneralBlazars}. In Section \ref{sec:method}, we describe the method for data preparation and \textit{Fermi}-LAT data analysis routines. In Section \ref{sec:taus} we present our findings in relation to the $\gamma$-ray flux variability timescales, and in Section \ref{sec:specvar} we explore the spectral shape during the flare period. This includes both the variation in the shape of the spectrum and an analysis of the high energy emission. In Section \ref{sec:hardness} we investigate whether or not an energy-dependence of the cooling of the emitting electron population exists, and we discuss the interpretation of the combined results in Section \ref{sec:discussion}. We summarise our conclusions in Section \ref{sec:conc}.

\section{Data Preparation and Source Modelling}
\label{sec:method}
The data used in this study were collected by the \textit{Fermi}-LAT (hereafter \textit{Fermi}). We consider photons detected in the energy range 0.1~$\leq E_{\gamma}\leq$~300~GeV, between mission elapse time (MET) 422409603 and 427248003. This corresponds to midnight on the 22nd May 2014 until midnight on 17th July 2014. The enhanced $\gamma$-ray emission from 3C~454.3 during this period was reported by \citet{ref:Buson2014} as the first flare was peaking, on 15th June 2014. For this reason, the flaring period is referred to as June 2014 in this paper. The region of interest (RoI) covers a radius of 15$\degree$ centred on 3C~454.3. A radius of 15$\degree$ was chosen to account for the point spread function (PSF) of the detected $\gamma$-rays. The PSF of a photon depends on the energy of the photon, due to differences in the photon scattering. The 68\% containment angle for $\gamma$-rays ranges from $\sim$6$\degree$ for 0.1~GeV photons to $\sim$0.2$\degree$ for 100~GeV photons \citep{ref:Atwood2009, ref:Ackermann2012}.

`Source'\footnote{`Source' class photons have an event class of 2 in the P7REP data. These have a high probability of being a photon \citep{ref:Ackermann2012}.} class photons were selected for analysis, and the instrument response function (IRF) used was P7REP$\_$SOURCE$\_$V15. As recommended by the P7REP data selection criteria, a zenith cut of 100$\degree$ was applied in order to exclude background photons from the Earth's atmosphere. The good time intervals were created by specifying that the LAT detector was at a rock angle of <~52$\degree$ and the filter expression `(DATA\_QUAL==1) \&\& (LAT\_CONFIG==1)' was satisfied.

In order to calculate the correct flux for each $\gamma$-ray source from the raw \textit{Fermi} data, a model was created containing the position and spectral definition for all of the point sources and diffuse emission in the RoI. The Galactic and extragalactic diffuse models used were gll\_iem\_v05\_rev1.fit and iso\_source\_v05.txt respectively. Both 3C~454.3 and neighbouring sources were modelled using the spectral definitions given in the 2-year 2FGL catalog\footnote{The \textit{Fermi}-LAT 4-year Point Source Catalog, 3FGL, was released during the writing of this paper. The 3FGL contains a greater number of $\gamma$-ray sources than the 2FGL \citep{ref:Acero2015}. However, the modelling and analysis routines performed in this paper ensure that accurate results are drawn from the photon data.}. The spectra of point sources in the region of interest are often modelled as power laws. The log parabola spectral shape of 3C~454.3, as well as of other blazars modelled in the RoI, is defined as:
\begin{equation}
\label{eqn:logparabola}
dN/dE = N_{0}(E/E_{b})^{-(\alpha+\beta(log(E/E_{b})))}
\end{equation}
where $dN/dE$ is the number of photons~cm$^{-2}$~s$^{-1}$~MeV$^{-1}$, $N_{0}$ is the normalisation of the energy spectrum, $E$ is the $\gamma$-ray photon energy and $E_{b}$ is the scaling factor of the energy spectrum. The $\alpha$ and $\beta$ (curvature) are spectral parameters. This spectral definition is taken from the \textit{Fermi} 2FGL catalog \citep{ref:Nolan2012}.

A binned analysis was run initially to find the spectral parameters that best describe each source during the period of interest. The MINUIT minimiser was used during all \textit{Fermi} gtlike optimisations. During this binned analysis, the spectral parameters of all of the sources in the RoI were free to vary. This ensured that the spectral parameters returned for each source provided an accurate representation of the spectral state of the source during the time period studied here.

\begin{figure*}
  \begin{minipage}{175mm}

\centering
\includegraphics[angle=0,width=.33\textwidth]{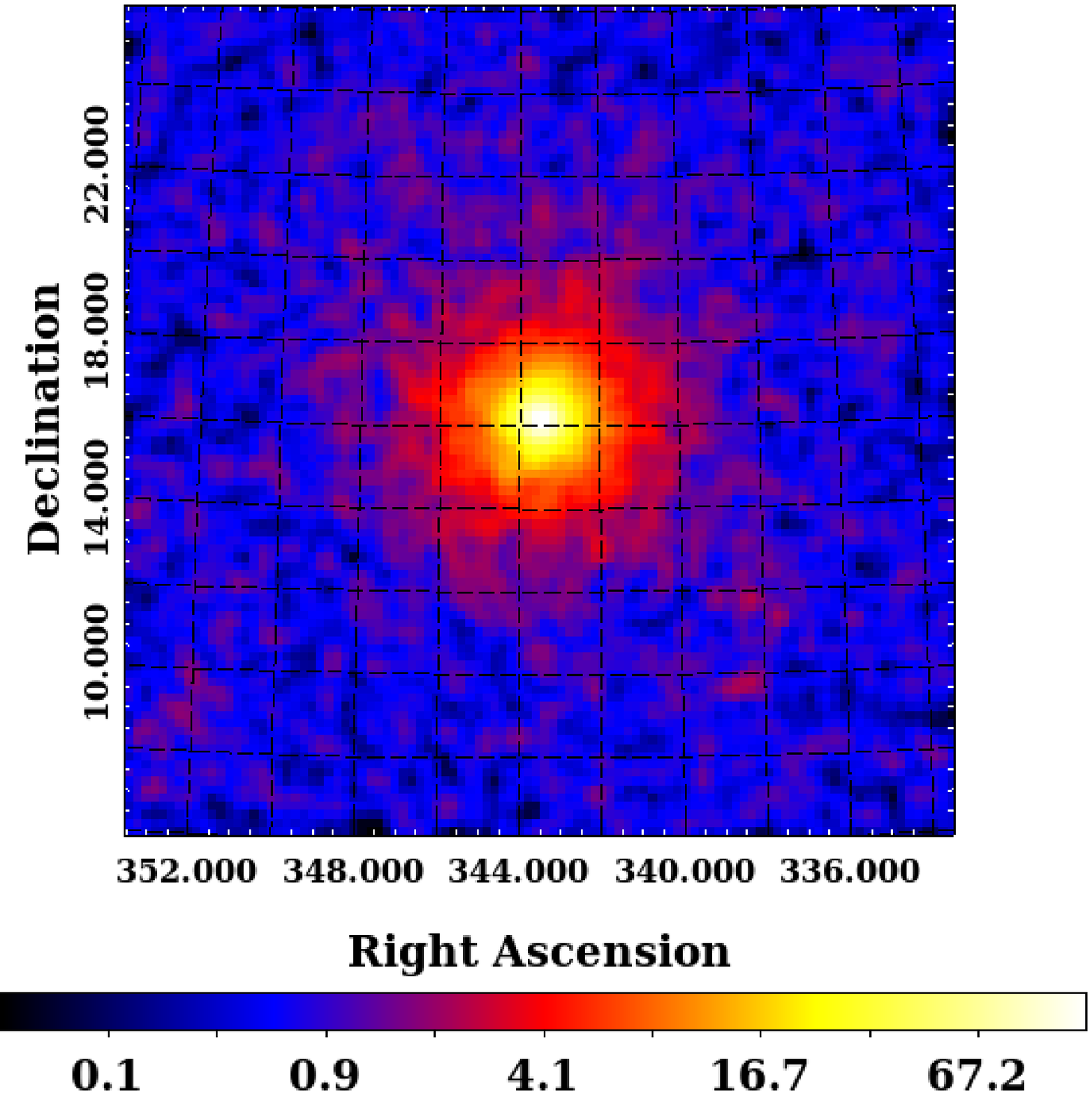}\hfill
\includegraphics[angle=0,width=.33\textwidth]{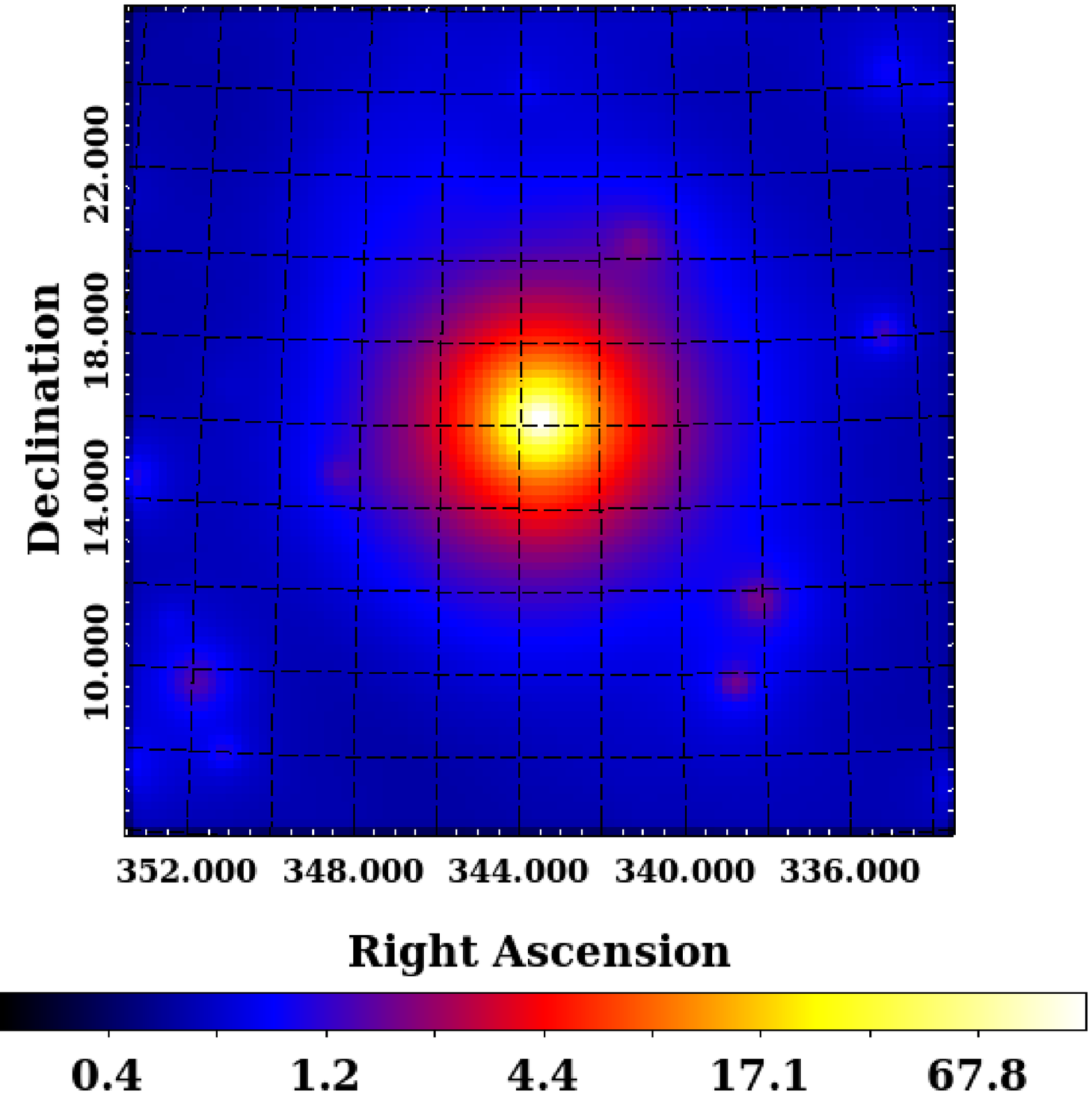}\hfill
\includegraphics[angle=0,width=.33\textwidth]{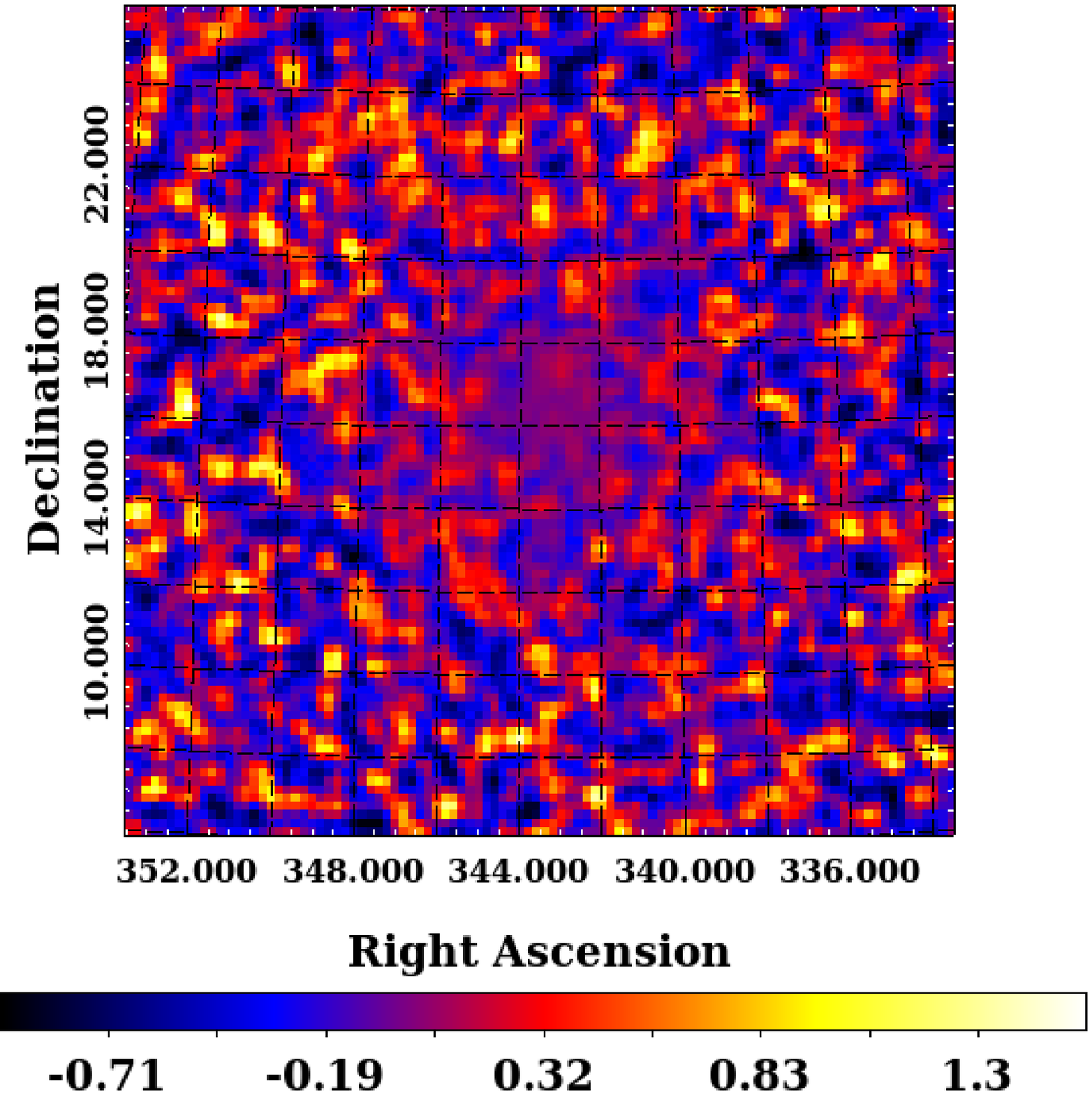}
\caption{20$\degree$ $\times$ 20$\degree$ observed (left), model (centre) and residuals (right) maps of the 0.1-300~GeV flux centred on 3C~454.3. The observed and model maps are in units of $\gamma$-ray counts, and the residuals map is in units of percentage. All maps are smoothed with a 2$\degree$ Gaussian and are at a scale of 0.2$\degree$/pixel.}\label{fig:cmaps}
  \end{minipage}
\end{figure*}

From the results of the binned analysis, we compared the observed $\gamma$-ray counts map with the model counts map of the RoI, created by the \textit{Fermi} gtmodel tool. This was done in order to assess whether or not any significant $\gamma$-ray sources existed in the RoI that had not been included in the 2FGL catalog. A residuals map was created by subtracting the model counts map from the observational counts map, and dividing by the model counts map. The observed map, model map and residuals map are shown in Fig.~\ref{fig:cmaps}. If any significant sources were found that were not present in the model, they could be added accordingly (e.g. \citealt{ref:Brown2015}). Creating these maps allowed us to be confident that all of the sources in the RoI were accounted for and that both the $\gamma$-ray sources and RoI were represented accurately across the time period under investigation. No significant sources were detected that had not already been included in the 2FGL catalog, so no additional sources were added to the model.

In order to study the $\gamma$-ray characteristics of 3C~454.3, the correct initial model of the RoI was used during the unbinned analyses. These analyses were then employed to create plots of flux and spectral parameters with time, presented in Sections \ref{sec:taus} - \ref{sec:hardness}.
For the unbinned analyses, spectral parameters $\alpha$, $\beta$ and $N_{0}$ of 3C~454.3 were input as the best-fitting parameters calculated from the binned analysis, but were free to vary during the gtlike fitting procedure. The spectra of all other sources except for the Galactic and extragalactic diffuse backgrounds were frozen at the best-fitting parameters returned by the binned analysis.

\section{Flux Variability Timescales}
\label{sec:taus}

Blazars are observed to be the most highly variable class of AGN. Strong $\gamma$-ray flux variability has been captured by the \textit{Fermi}-LAT as well as by ground-based instruments, when blazars exhibit an outburst well above their baseline emission. We use the term baseline to mean emission at a typical flux level for a given blazar. For 3C~454.3, this is $\sim$4.6~$\times$~$10^{-6}$~ph~cm$^{-2}$~s$^{-1}$ during the period of interest, from the average of our daily fluxes shown in Fig.~\ref{fig:alphaBetaTime}. The rapid variability during these flares allows us to probe more closely the physical processes occurring within the relativistic blazar jets.

The timescales on which we see $\gamma$-ray emission from an AGN vary allow us to constrain the size of the emission region. We assume that the $\gamma$-rays originate from within the relativistic jet, with the two main emission locations under consideration being the BLR and the narrow-line or MT region. The BLR is located close to the base of the jet and the SMBH, while the torus is further downstream \citep{ref:Urry1995}. The jet expands and widens with distance from the SMBH according to its opening angle (thought to be of the order of $\sim$0.1~rad, \citep{ref:GhiselliniTavecchio2009, ref:GeneralBlazars}), so that the cross-sectional diameter of the jet is smaller in the BLR than the MT. If we assume that the entire cross-section of the jet at a certain location is responsible for the emission, the light-crossing time and therefore the $\gamma$-ray flux doubling timescale will be smaller for a BLR origin.

Ground-based telescopes such as H.E.S.S. and MAGIC have measured $\gamma$-ray variability from blazars on extremely short timescales. Examples of these observed flux doubling times include $\sim$220~s \citep{ref:Aharonian2007}, $\sim$2~min \citep{ref:Albert2007}, and more recently <~5~min \citep{ref:Aleksic2014}. Previous studies using \textit{Fermi}, such as those done by \citet{ref:Brown1510} and \citet{ref:Saito2013}, have investigated flux doubling timescales using a minimum of 3 hour time bins in the unbinned \textit{Fermi} analysis. This duration is often chosen as the minimum because it is the time that the \textit{Fermi}-LAT takes to complete one full scan of the sky (2 orbits). In the case of the FSRQ PKS~1510-089, this revealed intrinsic doubling timescales, $\tau_{int}$, of $\tau_{int}=1.30$~$\pm$~0.12~h during the October 2011 flare period \citep{ref:Brown1510}. \citet{ref:Paliya2015} also found an observed variability timescale, $\tau$, of just $\tau=$~1.19~h at the $\sim$4$\sigma$ level for the FSRQ 3C~279 during a flare in March 2014. Another recent study of 3C~279 by \citet{ref:Hayashida2015} found a characteristic flux rising timescale of only $\tau=1.4$~$\pm$~0.8~h and a flux decay timescale of $\tau=0.68$~$\pm$~0.59~h, although the fitting errors are relatively large. However, we can probe down to the smallest timescales using good time interval (gti) time bins as described by \citet{ref:Foschini2011a, ref:Foschini2011b}. The gti time bins are uneven in length and are provided with the \textit{Fermi} raw photon data, with the binning being dependent on the instrument pointing direction \citep{ref:ItalianPKS1510}. The good time intervals are of the order of one orbit of the \textit{Fermi}-LAT, $\sim$90 minutes. This analysis technique enabled \citet{ref:ItalianPKS1510} to discover the fastest FSRQ $\gamma$-ray variation measured to date, during the same flare period of PKS 1510-089 as had been studied in 3 hour time bins. The $\gamma$-ray flux took just $\sim$20 minutes to double \citep{ref:ItalianPKS1510}. Doubling times of less than one hour enable us to put tight constraints on the size of the emission region.

\begin{figure*}
    \centerline{\includegraphics[width=21cm]{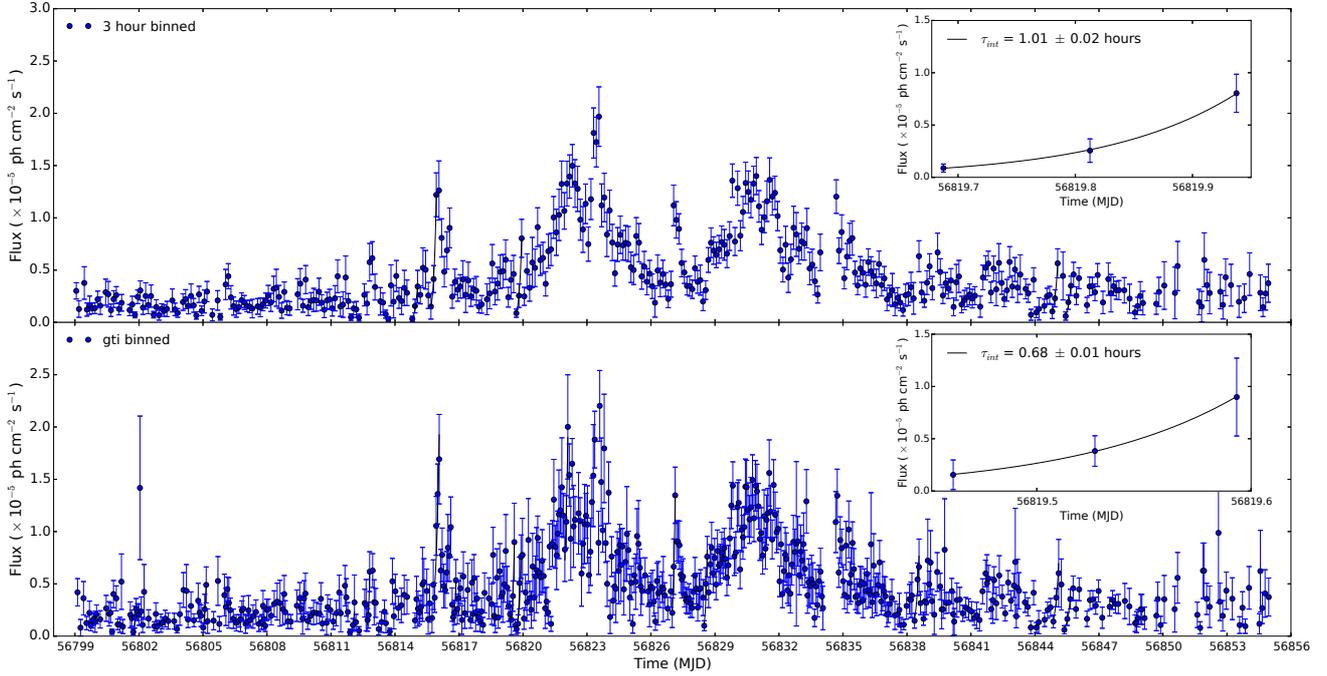}}
    \caption{The light curve of 3C~454.3 between MJD 56799 and 56855. Top: 3 hour binned. Bottom: gti binned. The horizontal error bars are not shown here, but are of unequal sizes for the gti binning. The best-fitting curves of equation \ref{eqn:taus} are also plotted (black lines). The insets show zoomed-in sections of the light curves, and the intrinsic doubling timescale is given in the legend. Only data points with $TS\geq10$ are shown.}\label{fig:3hourtausLC}
\end{figure*}

\begin{table*}
\begin{center}
\begin{tabular}{cccccc}
\hline
Start Time & End Time & $F(t_{0})$ & $F(t)$ & $\tau_{int}$ & Significance  \\
(MJD) & (MJD) & ($\times$ 10$^{-6}$ photons cm$^{-2}$ s$^{-1}$) & ($\times$ 10$^{-6}$ photons cm$^{-2}$ s$^{-1}$) & (hours) & ($\sigma$)  \\ \hline
56805.011 & 56805.150 & 0.82 $\pm$ 0.45 & 4.9 $\pm$ 1.7 & 0.71 $\pm$  0.09 & 8.04 \\
56819.461 & 56819.593 & 1.6 $\pm$ 1.4 & 9.0 $\pm$ 3.7 & 0.68 $\pm$  0.01 & 54.8 \\
56826.997 & 56827.134 & 2.2 $\pm$ 1.4 & 13 $\pm$ 3 & 0.72 $\pm$  0.08 & 8.86 \\
56844.863 & 56845.110 & 1.4 $\pm$ 0.6 & 6.0 $\pm$ 3.2 & 1.46 $\pm$  0.26 & 5.53 \\
56845.375 & 56845.655 & 0.61 $\pm$ 0.39 & 3.5 $\pm$ 0.9 & 1.47 $\pm$  0.11 & 13.6 \\
56819.697 & 56819.905 & 0.67 $\pm$ 0.50 & 7.6 $\pm$ 2.1 & 0.74 $\pm$  0.11 & 6.91 \\ \hline
56801.732 & 56801.864 & 0.40 $\pm$ 0.29 & 2.6 $\pm$ 1.0 & 0.67 $\pm$  0.19 & 3.54 \\
56811.893 & 56812.101 & 0.39 $\pm$ 0.28 & 3.2 $\pm$ 2.6 & 0.99 $\pm$  0.29 & 3.43 \\
56812.713 & 56812.814 & 1.5 $\pm$ 1.1 & 6.2 $\pm$ 1.9 & 0.54 $\pm$  0.15 & 3.52 \\
56816.731 & 56816.857 & 1.8 $\pm$ 1.4 & 5.3 $\pm$ 2.0 & 0.88 $\pm$  0.22 & 3.92 \\
56819.697 & 56819.836 & 0.67 $\pm$ 0.50 & 5.1 $\pm$ 2.5 & 0.56 $\pm$  0.15 & 3.74 \\
56821.962 & 56822.102 & 8.3 $\pm$ 3.1 & 20 $\pm$ 5 & 1.30 $\pm$  0.32 & 4.02 \\
56828.494 & 56828.634 & 1.0 $\pm$ 0.5 & 7.6 $\pm$ 1.9 & 0.66 $\pm$  0.20 & 3.22 \\
56838.355 & 56838.564 & 1.5 $\pm$ 0.7 & 6.6 $\pm$ 2.7 & 1.22 $\pm$  0.32 & 3.82 \\
56815.865 & 56816.074 & 3.7 $\pm$ 1.4 & 17 $\pm$ 4 & 1.34 $\pm$  0.37 & 3.66 \\ \hline
\end{tabular}
\end{center}
\caption{$\gamma$-ray flux intrinsic doubling timescales and their significance, from the gti unbinned analysis. Above the horizontal line, the timescales are significant at the $\geq 5\sigma$ level.}\label{tab:gti_taus}
\end{table*}
\begin{table*}
\begin{center}
\begin{tabular}{cccccc}
\hline
Start Time & End Time & $F(t_{0})$ & $F(t)$ & $\tau_{int}$ & Significance \\
(MJD) & (MJD) & ($\times$ 10$^{-6}$ photons cm$^{-2}$ s$^{-1}$) & ($\times$ 10$^{-6}$ photons cm$^{-2}$ s$^{-1}$) & (hours) & ($\sigma$)  \\ \hline
56814.813 & 56815.063 & 0.34 $\pm$ 0.25 & 3.4 $\pm$ 1.5 & 1.01 $\pm$  0.20 & 5.01 \\
56815.688 & 56815.938 & 1.5 $\pm$ 1.2 & 12 $\pm$ 2 & 1.05 $\pm$  0.02 & 52.6 \\
56819.688 & 56819.938 & 0.88 $\pm$ 0.39 & 8 $\pm$ 2 & 1.01 $\pm$  0.02 & 56.4 \\
56844.813 & 56845.063 & 1.1 $\pm$ 0.7 & 5.7 $\pm$ 1.4 & 1.30 $\pm$  0.05 & 25.6 \\
56845.438 & 56845.688 & 0.61 $\pm$ 0.39 & 3.5 $\pm$ 0.9 & 1.39 $\pm$  0.27 & 5.17 \\
    \hline
\end{tabular}
\end{center}
\caption{$\gamma$-ray flux intrinsic doubling timescales and their significance, from the 3 hour unbinned analysis.}\label{tab:3hour_taus}
\end{table*}

The flux doubling timescales of the $\gamma$-ray data are calculated using:
\begin{equation}
\label{eqn:taus}
F(t) = F(t_{0})2^{(\tau^{-1}(t-t_{0}))}
\end{equation}
where $F(t)$ and $F(t_{0})$ are the flux at times \textit{t} and \textit{$t_{0}$} respectively and $\tau$ is the observed flux doubling timescale. We can use this observed timescale to calculate the intrinsic doubling timescale by taking into account the redshift of 3C~454.3, z = 0.859 \citep{ref:Jackson1991}. A least-squares routine was used to calculate the parameters that provide a best-fit solution to equation \ref{eqn:taus} for flare rise and fall subsets of our data. Examples of the resulting curves are shown as the insets in Fig.~\ref{fig:3hourtausLC}. This fitting was done for both 3 hour and gti binned data. Tables \ref{tab:gti_taus} and \ref{tab:3hour_taus} present the intrinsic doubling timescales of the $\gamma$-ray flux for several time intervals between MJD 56799 and 56855. Only data points with a gtlike analysis test statistic\footnote{The test statistic is defined as $TS=-2ln(L_{0}/L_{1})$, where $L_{0}$ is the maximum likelihood value for a model when the source is not included, and $L_{1}$ is the maximum likelihood value for a model with the source included at the specified location \citep{ref:Mattox1996}.} $TS\geq10$ ($\sim$3$\sigma$) are considered. Doubling timescales that are $\leq1.5$~h with a significance of $\geq3\sigma$ are shown. The errors given on the timescales in Tables \ref{tab:gti_taus} and \ref{tab:3hour_taus} are one standard deviation, $\sigma$. The significance of a doubling timescale in terms of $\sigma$ is defined as how many standard deviations $\tau_{int}$ is from zero. Interestingly, no flux halving timescales that fit these criteria were found. The fastest doubling timescale that we discover is $\tau_{int}$ = 0.68 $\pm$ 0.01~h, between MJD 56819.461 and 56819.593.

From Table \ref{tab:gti_taus}, we identify four occasions on which the flux doubles in less than one hour at a significance level >~5$\sigma$. The size of the emission region can be constrained using:
\begin{equation}
\label{eqn:emissionSize}
R \leq c\delta\tau_{int}
\end{equation}
where $R$ is the diameter of the emission region, $c$ is the speed of light, $\delta$ is the Doppler factor of the jet and $\tau_{int}~=~\tau/(1+z)$. These sub-hour flux doubling times therefore imply that the size of the emission region is relatively small. Taking equation \ref{eqn:emissionSize} and $\tau_{int}=0.68$ h, we are able to constrain the size of the emission region to $R\delta^{-1}~\leq~2.38~\times~10^{-5}$~pc.

We can use the intrinsic doubling timescales in Table \ref{tab:gti_taus} to calculate the required Doppler factor of the jet, if we assume a minimum size for the emission region. The mass of the SMBH, \Mbh, of 3C~454.3 is $(0.5~-~4.6)$~$\times$~$10^{9}$\Msol \citep{ref:BHMass, ref:Bonnoli2011}. The corresponding range for the Schwarzschild radius of the SMBH, $R_{S}$, is $(0.48 - 4.40)$~$\times$~$10^{-4}$~pc. This can be taken as the smallest cross-sectional radius of the jet, provided that the jet doesn't re-collimate downstream. We might assume that the Schwarzschild radius is therefore the minimum radius of the $\gamma$-ray emission region. This is a conservative assumption, as the emission region could also take the form of a small blob within the jet. The range of minimum Doppler factor required for a flux doubling timescale $\tau_{int}$ = 0.68~h is therefore $\delta_{min}$ = 4.03 - 37.03, from equation \ref{eqn:emissionSize}. However, we note that a value of $\delta$ as low as $\delta$~=~4.03 is not consistent with previous measurements of $\delta$ \citep{ref:Jorstad2005, ref:Ackermann2010, ref:Abdo2011} for 3C~454.3. \citet{ref:Jorstad2005} used very long baseline interferometry (VLBI) observations of 3C~454.3 to derive jet Doppler factors $\sim$14-30. We will discuss relevant emission region models for the June 2014 flares in Section~\ref{sec:discussion}.

The size of the emission region compared with the jet may be dependent on factors such as the Doppler factor of the jet and the geometry of the jet, as described above. If we take $\delta$ = 25 for the sake of argument, which is consistent with the literature \citep{ref:Jorstad2005} and with the above calculations, we find $R\leq5.95 \times 10^{-4}$~pc for $\tau_{int}$ = 0.68~h.

It should be highlighted that constraining the size of the emission region does not allow us to locate the emission region. However, once the location of the emission region is inferred, it will be interesting to compare the size of the emission region with the size of the jet at that distance.

\section{Spectral Shape and Photon-Photon Pair Production}
\label{sec:specvar}
High energy photons, such as $\gamma$-rays at $E_{\gamma}$~>~10~GeV, can be absorbed by lower energy optical and UV photons. This leads to photon-photon pair production ($\gamma\gamma\rightarrow e^{+}e^{-}$). FSRQs such as 3C~454.3 are very bright in their innermost regions, near to the SMBH and the accretion disk. Several studies have shown that $\gamma$-rays emitted here can be absorbed by the lower energy ambient photons in these inner regions \citep{ref:Donea2003, ref:Lui2006}. In this way, high energy $\gamma$-rays originating close to the base of the jet are absorbed before they are able to escape the BLR. The photon field external to the jet is comprised of photons from the accretion disk, the reprocessed emission in the BLR or MT, and thermal radiation from the corona close to the SMBH \citep{ref:IRphots, ref:Sikora2002, ref:GhiselliniTavecchio2009}. The density of this photon field at a certain point along the jet is dependent on the distance to the SMBH and the accretion disk luminosity \citep{ref:GhiselliniTavecchio2009, ref:Sikora2009, ref:Pacciani2014}. The photon density in the MT is therefore much lower than inside the BLR, greatly increasing the likelihood of pair production in the BLR compared to the MT. This photon attenuation manifests itself as a high energy cut-off in the $\gamma$-ray spectrum, such that the shape of the spectrum may be better described as a log parabola than a power law. As the MT is not as opaque to high energy photons as the BLR, we would not expect to observe a cut-off due to attenuation if the $\gamma$-rays are being produced here. It has been suggested that the spectral shape of 3C~454.3 can be fitted well by a log parabola or a broken power law, where we see $\gamma$-rays being emitted from the base of the jet (e.g. \citealt{ref:Ackermann2010, ref:Poutanen2010, ref:Harris2012}). It has also been suggested that specific GeV breaks that are present in the spectra of some AGN could be arising due to pair production of $\gamma$-rays with the He Lyman recombination continuum, again in the BLR \citep{ref:Poutanen2010}. However, \citet{ref:Harris2012} found that the location of these spectral breaks was inconsistent with the absorption model proposed. A log parabola spectral shape might also arise from a curved energy distribution of the emitting electrons \citep{ref:Dermer2015}.

\subsection{Spectral Variation}
As discussed in Section \ref{sec:method}, the spectral shape of 3C~454.3 is modelled by a log parabola, equation \ref{eqn:logparabola}. The log parabola is the spectral shape used to describe 3C~454.3 in the first two \textit{Fermi}-LAT catalogues (1FGL, 2FGL). The parameter $\alpha$ dictates the slope of the spectrum and is therefore a measure of the hardness of a spectrum, with a shallower slope indicating relatively more high energy emission. The amount of curvature in the spectrum is described by $\beta$. This curvature leads to a cut-off in the flux at higher energies, with a larger curvature giving a sharper cutoff.

\begin{figure*}
    \centerline{\includegraphics[width=21cm]{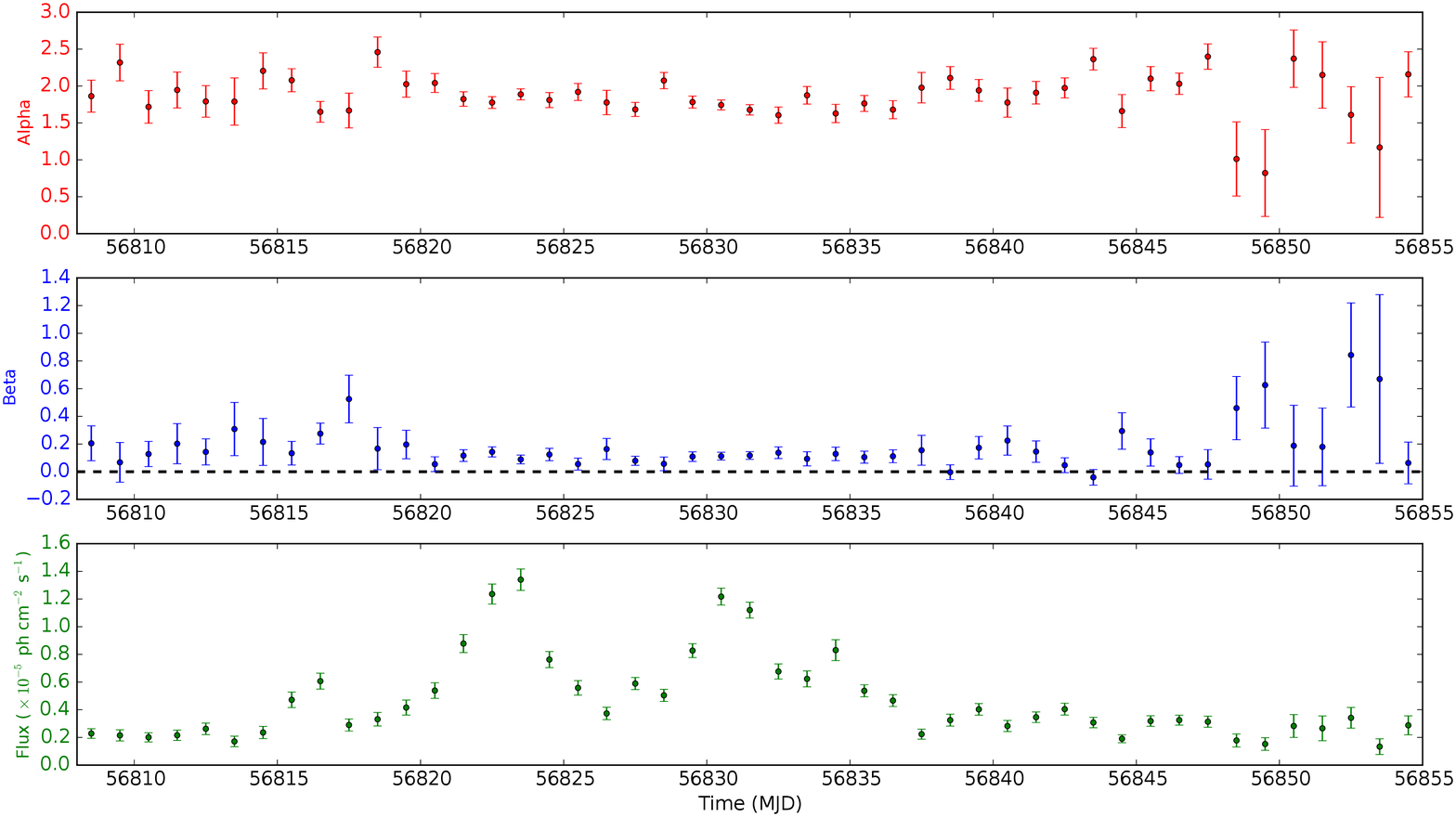}}
    \caption{Top: $\alpha$ as a function of time. Middle: The variation of $\beta$ with time. The dashed line is at $\beta$=0. Bottom: The $\gamma$-ray light curve. All three plots are binned daily. No strong trend of $\alpha$ and $\beta$ with flux is identified. The curvature during the flares is lower and less turbulent than during the baseline emission. All of the data points have a $TS\geq25$.}\label{fig:alphaBetaTime}
\end{figure*}

\begin{figure}
    \centerline{\includegraphics[width=10.1cm]{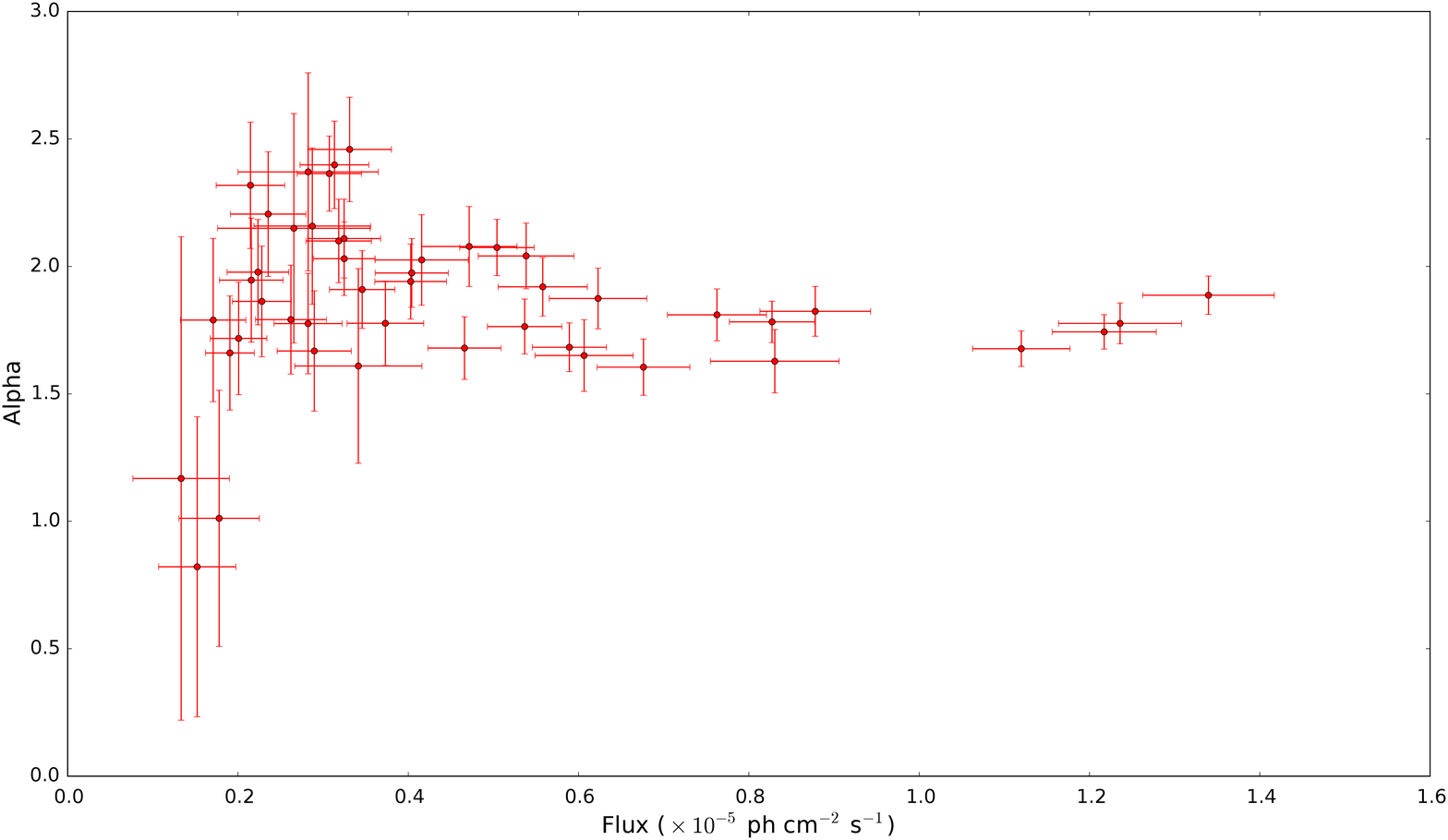}}
    \caption{$\alpha$ as a function of flux, binned daily. All of the data points have a $TS\geq25$.}\label{fig:alphaFlux}
\end{figure}
\begin{figure}
    \centerline{\includegraphics[width=10.1cm]{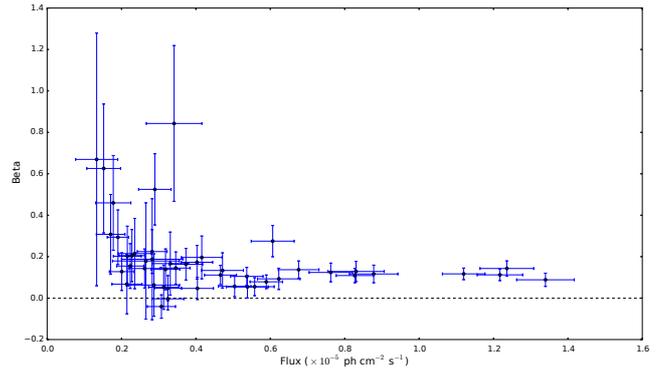}}
    \caption{$\beta$ as a function of flux, binned daily. The dashed line is at $\beta$ = 0. All of the data points have a $TS\geq25$.}\label{fig:betaFlux}
\end{figure}

We consider first the presence of spectral variation with time. Changes in the $\alpha$ and $\beta$ parameters of a log parabola indicate that the spectral shape of the $\gamma$-rays is changing. One possible reason for this is that the dominant location of $\gamma$-ray emission is changing, particularly during a flare. A change in the spectral shape during a flare gives us information on both the point-of-origin of the flare, and whether the emission location is different to that of the baseline $\gamma$-ray emission. Spectral changes with increasing flux are of interest as they could indicate injection into the high energy part of the spectrum, or a decrease in high energy attenuation.

The data in Fig.~\ref{fig:alphaBetaTime} are binned daily in order to observe trends in the spectral parameters without sacrificing the statistics. All of the data have a $TS\geq25$ from the gtlike analysis. We can see from Fig.~\ref{fig:alphaBetaTime} that both $\alpha$ and $\beta$ vary with time across June 2014, which tells us that the spectral shape of 3C~454.3 is changing across the flare period.

There doesn't appear to be strong evidence for the spectrum becoming harder as the FSRQ gets brighter, or for a correlation between curvature and flux. In order to investigate this further, it is also useful to look at $\alpha$ and $\beta$ as a function of flux. Figs \ref{fig:alphaFlux} and  \ref{fig:betaFlux} are binned daily, and show that the relationship between both $\alpha$ and $\beta$ with flux becomes flatter at higher flux. At low flux, we see the lowest values of $\alpha$ in Fig.~\ref{fig:alphaFlux}, although there are large error bars on these points. We also see the highest curvature at low flux in Fig.~\ref{fig:betaFlux}, again with large error bars. These days correspond to MJD 56848 onwards, as can be seen from Fig.~\ref{fig:alphaBetaTime}. This is $\sim$10 days after the second flare has finished, and the flux is $\sim$5 times less than during the peak of the flares. The large errors on $\alpha$ and $\beta$ at this time are likely to be due to poor photon statistics; there are also relatively large errors on the flux here, seen in Fig.~\ref{fig:alphaBetaTime}.

We now look specifically at how $\alpha$ is changing during the flare period. Fig.~\ref{fig:alphaFlux} does not show a strong correlation between the hardness of the source and flux. When the flux reaches above $F=0.6\, \times10^{-5}$~ph~cm$^{-2}$~s$^{-1}$, $\alpha$ remains between 1.5 and 2 and the distribution is relatively flat. A spectral index of <~2, or $\alpha$~<~2 in the case of a log parabola, is generally considered to be a hard spectrum. At lower flux there is a greater range of $\alpha$, but a slightly negative correlation can be found, suggesting a harder-when-brighter behaviour. This could either be due to the energy of the emitting electron population, or an indication that the high energy emission suffers less from absorption during the flares.

We next examine how $\beta$ is changing. When $\beta$ = 0, the spectral shape of a log parabola is equivalent to a power law, as we no longer have any spectral curvature. A low value of $\beta$ therefore strongly suggests a power law spectral shape if the error on $\beta$ is small. When $\beta$ is larger, such as on MJD 56849 in Fig.~\ref{fig:alphaBetaTime}, the spectral shape takes the form of a log parabola. However, if the error on $\beta$ is also large, we cannot conclude that strong curvature is present. Despite the large errors on $\beta$ either side of the flares in Fig.~\ref{fig:alphaBetaTime}, variation in $\beta$ can be identified between MJD 56808 and 56855. The curvature between MJD 56820-38, during the flares, appears to be lower than during the baseline emission. This is supported by the results of a least-squared analysis performed to find the best-fitting constant value of $\beta$. The analysis was done for the entire period between MJD 56808 and 56855, and also between MJD 56818 and 56038, across the flares. A higher value of $\beta$ was returned across the entire period compared to during the flares. The reduced chi-squared values for the fits are $\chi_{red}^{2}=$~2.86 and $\chi_{red}^{2}=$~0.65 respectively, demonstrating that a lower value and flatter distribution of $\beta$ are better fits to the data during the flares. Current evidence for an unambiguous variation of the spectral curvature between MJD 56808 and 56855 is therefore suggestive, although not yet compelling. Whilst other possible explanations exist, this could be interpreted as evidence that the flaring emission region is at a different location to that of the baseline emission. Better statistics would be required to probe the spectral shape either side of the flares and explore the idea of multiple emissions regions further.

Fig.~\ref{fig:betaFlux} shows no strong trend in curvature, due to the errors on $\beta$ being large at low flux. It can however be seen that there is curvature on the majority of days between MJD 56808 and 56855. Approximately ten days during this period are consistent with $\beta$ = 0, but there is certainly evidence for spectral curvature across the flux range. There is a trend towards larger curvature at lower flux, even with the larger error bars being taken into account. At higher flux, above $F = 0.6$ $\times$ $10^{-5}$~ph~cm$^{-2}$~s$^{-1}$, the distribution of $\beta$ becomes much flatter and $\beta$ is not consistent with 0. We note that a process other than pair production in the BLR could be responsible for the small amount of curvature that we see during the flares, and this will be discussed in Section \ref{sec:vhe}.

We need to assess whether or not the trend we see for $\beta$ decreasing with flux is solely due to poor photon statistics at the low flux end. The trend of decreasing curvature that we see in Fig.~\ref{fig:betaFlux} could be due to the fact that a curved spectrum is more easily fitted when there are large error bars on the flux, or it could represent a real change in the emission characteristics. To determine which interpretation is most likely, we compare Fig.~\ref{fig:betaFlux} to fig.~16 of \mbox{\citet{ref:Ackermann2015}}, from the third catalog of AGN detected by the \textit{Fermi}-LAT (3LAC). The 3LAC plot shows $\beta$ against flux for four years of observations on FSRQs and BL Lac objects. Theirs is a phenomenological study with a large data set, but it shows the same shape and trend of $\beta$ with flux as in the daily binned Fig.\mbox{~\ref{fig:betaFlux}}. The 3LAC data is also less limited by statistics than \mbox{Fig.~\ref{fig:betaFlux}}. Nonetheless, further work is needed to make firm conclusions, but the similarity in trend might indicate that the property of a smaller curvature at larger flux is not simply due to poor statistics in the case of these flares.

\subsection{VHE Emission}
\label{sec:vhe}
VHE emission is defined as observed emission from a source at $E_{\gamma}\geq100$~GeV. At present, there are only 5 FSRQs that have been observed as VHE emitters\footnote{http://tevcat.uchicago.edu/ (accessed on 15/05/15)}, of which 3C~454.3 is not one. This is most likely due to the attenuation at high energy for the case of a BLR origin, which is classically assumed to be the location of the emission region. Observations by \citet{ref:Abdo2009} showed that the majority of blazars that emit in the TeV energy range have a hard photon spectrum, with a photon index < 2. As 3C~454.3 displays a similarly hard spectrum throughout the flare period, an unbinned analysis was done to calculate the significance of $E_{\gamma}\geq100$~GeV emission between MJD 56808 and 56855. No significant emission at $E_{\gamma}\geq100$~GeV was found. 3C~454.3 is therefore not a VHE emitter during this period, despite the hardness of the $\gamma$-ray spectrum seen in Fig.~\ref{fig:alphaFlux}. A lack of VHE emission could be due to a high energy cut-off, caused by the curvature of the spectrum that we see in Fig.~\ref{fig:betaFlux}. This might suggest that either the emitting electrons are not energetic enough to produce VHE $\gamma$-rays, or that there is a mechanism for high energy $\gamma$-ray attenuation taking place during flaring episodes.

We next probe the significance of $E_{\gamma}\geq$ 20~GeV high energy emission during the flares. We bin the data into 5 day periods between MJD 56810 and 56845, so that the statistics are good enough to detect the presence of high energy emission both during the flares and either side. The 5 day binned $\alpha$ values confirm that $\alpha$ remains below 2 across this period, and we see that $E_{\gamma}\geq$ 20~GeV emission is only significant during MJD 56825-30 and MJD 56830-35. The significance of the emission is > 5$\sigma$ and > 8$\sigma$ respectively. The 5 day binned values of $\alpha$ are consistent within error between MJD 56810-45, except for over MJD 56830-35. Here, we see $\alpha$ reach a minimum, at a value of $\alpha=1.72~\pm~0.04$. This suggests a trend towards spectral hardening at the peak of the second flare.

\begin{figure*}
     \centering
   \centerline{\includegraphics[width=20cm]{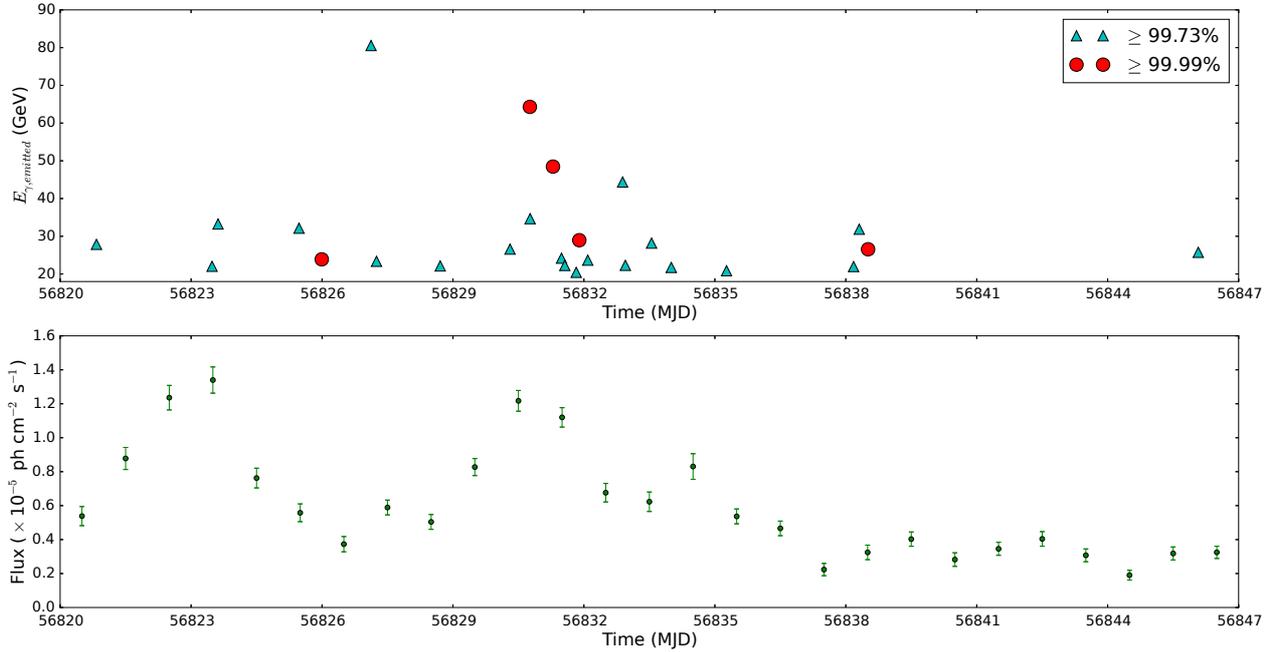}}
    \caption{The emitted energies of the individual high energy photons detected by \textit{Fermi} over the flare period, as a function of time. Only photons with $E_{\gamma, emitted}\geq$~20~GeV and a probability of originating from 3C~454.3 of $\geq$~99.73\% are shown.}\label{fig:PhotonEnergies}
\end{figure*}

\begin{table*}
\begin{center}
\begin{tabular}{ccc}
\hline
Emitted Energy (GeV) & Significance ($\sigma$) & Flux ($\times$ 10$^{-6}$ photons cm$^{-2}$ s$^{-1}$)\\ \hline
$\geq$ 35 & 9.8 & 2.57 $\pm$ 0.88 \\
$\geq$ 50 & 6.8 & 1.14 $\pm$ 0.57 \\
    \hline
\end{tabular}
\end{center}
\caption{The flux and significance for $E_{\gamma, emitted}\geq$ 35~GeV and $E_{\gamma, emitted}\geq$ 50~GeV emission, between MJD 56808 and 56855. A single time bin was used over this period.}\label{tab:HighESig}
\end{table*}

In order to quantify the position of the emission region, we study the optical depth of different \textit{emitted} photon energies with distance from the SMBH, both within the BLR and beyond. The expansion of the Universe means that \textit{Fermi} will detect $\gamma$-rays at $E_{\gamma}=E_{\gamma, emitted}/(1+z)$. The photon optical depth is a measure of how opaque a region is, in terms of how far a $\gamma$-ray can travel before being absorbed through $\gamma$-$\gamma$ pair production in this case. The intensity of the external photon field at a certain point along the jet dictates the optical depth of $\gamma$-rays of energy $\epsilon$, $\tau_{\gamma\gamma}(\epsilon)$. This is why the optical depth outside the BLR is much lower than inside \citep{ref:Lui2006}.

\citet{ref:Pacciani2014} interpolated the work of \citet{ref:Lui2006} in order to calculate optical depths for the BLR region of 3C~454.3. They found $\tau_{\gamma\gamma}(\epsilon)=2.8$ at 35~GeV and $\tau_{\gamma\gamma}(\epsilon)=4.0$ at 50~GeV, for $\gamma$-rays emitted at the mid-point of a spherical BLR shell. These are the \textit{emitted} $\gamma$-ray energies at the source. In order to interpret these optical depths in the context of our \textit{Fermi} data, we need to run analyses on the \textit{observed} energies that correspond to \textit{emitted} energies of 35~GeV and 50~GeV. The optical depths presented by \citet{ref:Pacciani2014} give a clear indication that we do not expect to observe significant $\gamma$-ray emission of $E_{\gamma, emitted}\geq$ 35~GeV and $E_{\gamma, emitted}\geq$~50~GeV if the $\gamma$-rays are being emitted at the mid-point of the BLR. Using the \textit{Fermi} tools to calculate the flux between MJD 56808 and 56855 for $\textit{emitted}$ energies\footnote{This corresponds to \textit{observed} energies, as detected by \textit{Fermi}, of $E_{\gamma}\geq$~$35/(1+z)$~GeV and $E_{\gamma}\geq$~$50/(1+z)$~GeV respectively.} $E_{\gamma, emitted}\geq$~35~GeV and $E_{\gamma, emitted}\geq$~50~GeV gives the fluxes shown in Table \ref{tab:HighESig} at significances of 9.8$\sigma$ and 6.8$\sigma$ respectively. These high energy fluxes are both significant, meaning that the emission region during these flares is extremely unlikely to be located in the middle of the BLR, due to the high opacity at these energies. The optical depth of the $\gamma$-rays will decrease with distance towards the outer edge of the BLR and beyond, so it's much more likely that the emission region is towards this outer edge. However, the existence of an axion-like particle (ALP) that could facilitate the path of photons through the BLR should also be considered. This mechanism has been postulated in order to explain the detection of VHE emission from distant sources \citep{ref:Csáki2003, ref:Harris2014}.

Taking the distance to the outer edge of the BLR, $R_{BLR}^{out}$, to be $\sim$3.8 times larger than the inner radius of the BLR, $R_{BLR}$ \mbox{\citep{ref:GhiselliniTavecchio2009, ref:Pacciani2014}}, we use the optical depth relations given in \citet{ref:Tavecchio2013} to assess at what distance along the jet the optical depth reaches a value of $\tau_{\gamma\gamma}(\epsilon)=1$ for $E_{\gamma, emitted}=35$~GeV and $E_{\gamma, emitted}=50$~GeV photons. We find that in both cases, the optical depth does not decrease to a value of 1 until the emission region is outside the BLR. In the case of $E_{\gamma, emitted}=35$~GeV photons, $\tau_{\gamma\gamma}(\epsilon)=1$ at $\sim$4.0~$\times$ $R_{BLR}$, equivalent to $\sim$1.1~$\times$ $R_{BLR}^{out}$. For $E_{\gamma, emitted}=50$~GeV photons, $\tau_{\gamma\gamma}(\epsilon)=1$ at $\sim$4.8~$\times$ $R_{BLR}$ or $\sim$1.3~$\times$ $R_{BLR}^{out}$. These results suggest that the emission region of these high energy $\gamma$-rays is located outside the BLR.

\begin{figure*}
     \centering
   \centerline{\includegraphics[width=20cm]{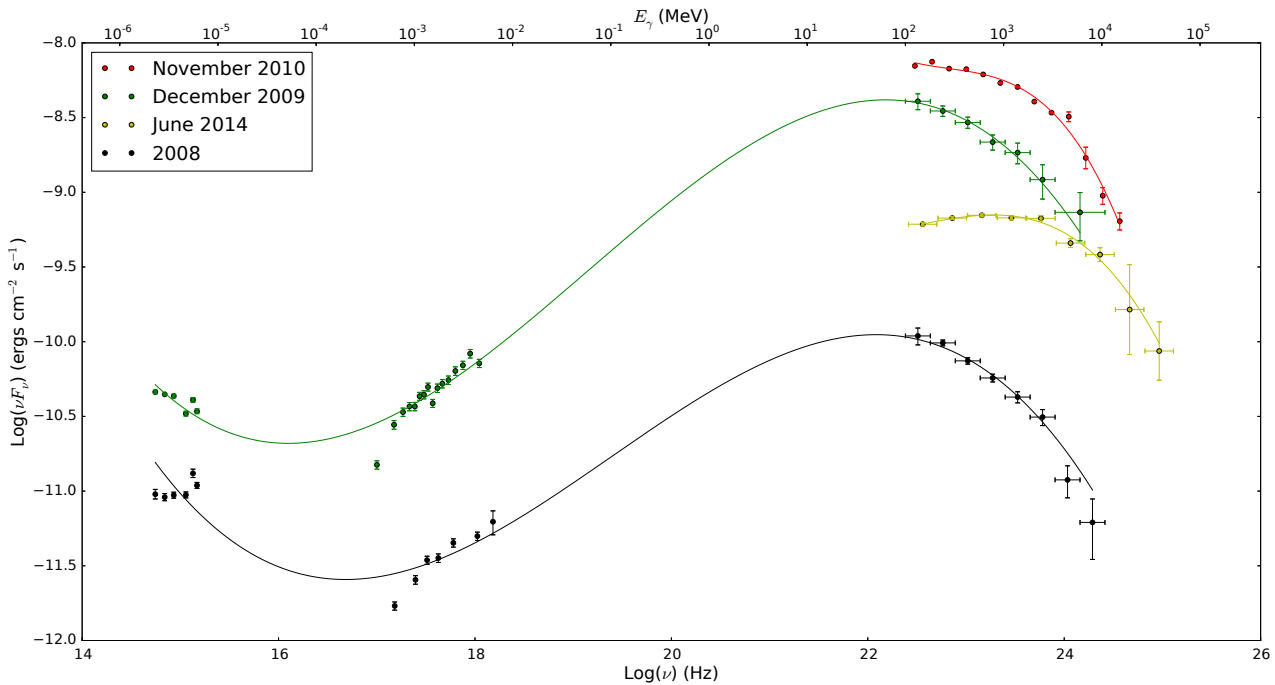}}
    \caption{The high energy SED of 3C~454.3 during December 2008 - May 2009 (black), December 2009 (green), November 2010 (red) and June 2014 (yellow). The 2008 and 2009 data are from \citet{ref:Bonnoli2011} and the 2010 data are from \citet{ref:Abdo2011}. The June 2014 spectrum uses the data analysed in this paper. In all cases, the $\gamma$-ray data were observed by the \textit{Fermi}-LAT. The optical-UV and x-ray data were observed by the \textit{Swift} satellite, using the Ultraviolet-Optical Telescope and the X-Ray Telescope respectively. The curves are best-fit third degree polynomials that have been calculated using a least-squares approach.}\label{fig:SED_June_3C454}
\end{figure*}

In order to dissect the high energy emission further, the \textit{Fermi} tool gtsrcprob was used to calculate the probability of each detected $E_{\gamma, emitted}\geq$~20~GeV photon having been emitted by 3C~454.3. Only photons within a radius of 0.1$\degree$ around 3C~454.3 were selected for analysis. Fig. \ref{fig:PhotonEnergies} shows the individual \textit{emitted} energies of the photons that were given a $\geq$~99.7\% probability of originating from 3C~454.3, and the time at which they were detected. The largest number of these high energy photons are emitted between MJD 56824 and 56835, corresponding to the fall of the first flare until a few days after the peak of the second flare. There are 26 photons at $E_{\gamma, emitted}\geq$~20~GeV in total, and the highest energy photon that is detected has an energy $E_{\gamma, emitted}=$~80~GeV. Interestingly, the highest energy photons are detected between MJD 56827 and 56833, coinciding closely with the second flare. This analysis of the high energy photons emitted by 3C~454.3 supports the result that there is a spectral hardening between MJD 56830-35, and that significant high energy emission is emitted across the flare.

If the emission region is located at r~$\sim$1.3~$\times$ $R_{BLR}^{out}$, a different model for the spectral curvature seen in Fig.~\ref{fig:betaFlux} than pair production within the BLR is required. \citet{ref:Pacciani2014} and \citet{ref:Tavecchio2008} studied the effect of the Klein-Nishina (KN) suppression. The KN regime of IC scattering that occurs in the BLR results in a curvature of the $\gamma$-ray spectrum at high energies. \citet{ref:Pacciani2014} show in fig.~7 that this suppression alone causes a curvature in the spectrum until at least $\sim$8 $\times$ $R_{BLR}$. This is consistent with the distance constraint on the emission region derived from the June 2014 data, based on the presence of high energy emission.

\subsection{Spectral Energy Distribution}

Fig. \ref{fig:SED_June_3C454} shows the high energy SED of 3C~454.3 at several different epochs, in order to compare the June 2014 flare with previous $\gamma$-ray flares. Best-fit SED curves were calculated for each epoch using a third degree polynomial, taking a least-squares approach. Fig. \ref{fig:SED_June_3C454} illustrates that although the June 2014 flare has a lower peak $\gamma$-ray flux than previous bright $\gamma$-ray flares, it is certainly significantly brighter than the quiescent state of 3C~454.3. However, the most notable feature in Fig. \ref{fig:SED_June_3C454} is the relative position of the $\gamma$-ray peak frequency. It can be seen for the 2008, 2009 and 2010 SEDs that the peak frequency in the \textit{Fermi}-LAT energy range corresponds to a photon energy $E_{\gamma}\sim$~150~MeV. For the June 2014 flare on the other hand, the peak is shifted to higher energies and lies between the $E_{\gamma}=$~600~MeV and $E_{\gamma}=$~1200~MeV energy bins. This is emphasised by the spectral curvature of the 2014 SED between $E_{\gamma}=$~150~MeV and $E_{\gamma}=$~2400~MeV, compared with the other observations. It can be seen that the $\gamma$-ray data for June 2014 illustrate a rise, peak and fall of the IC SED component. For the previous epochs, as the IC peak occurs at lower energies, the $\gamma$-ray data show only the fall of the IC component and may not contain the peak. This indicates that there is relatively more high energy emission in the June 2014 flare, compared with previous flares of 3C~454.3 and the quiescent state observed in 2008.

It has been discussed by \citet{ref:Sol2013} that $\gamma$-ray flares can be observed for a number of different reasons. These include an injection of particles into the jet, an increase in energy of the particles due to acceleration, and procession and beaming effects (e.g. \citealt{ref:Melrose2009, ref:Sironi2009, ref:Katarzynski2010}). It has been suggested that the movement of the synchrotron self-Compton (SSC) IC $\gamma$-ray peak of an SED with time can give insight into the mechanism that is causing the flare \citep{ref:Sol2013}. As can be seen in Fig. \ref{fig:SED_June_3C454}, the energy range over which the spectrum of 3C~454.3 is constructed in June 2014 is limited, and to probe the physical processes that are dominating the IC peak would require an extended SED before and after the flare. Even so, if the 2008 SED is an accurate representation of 3C~454.3 during a quiescent state then the shift in peak frequency may be noteworthy. From \citet{ref:Sol2013}, the results of Fig. \ref{fig:SED_June_3C454} suggest that an acceleration of the emitting particles may best describe the shift of the SED peak between the quiescent state and the June 2014 flare. However, as this modelling is based on a SSC model, the results may be more relevant to BL Lacs than FSRQs.

\section{Energy-dependent Cooling}
\label{sec:hardness}

\begin{figure*}
    \centerline{\includegraphics[width=21cm]{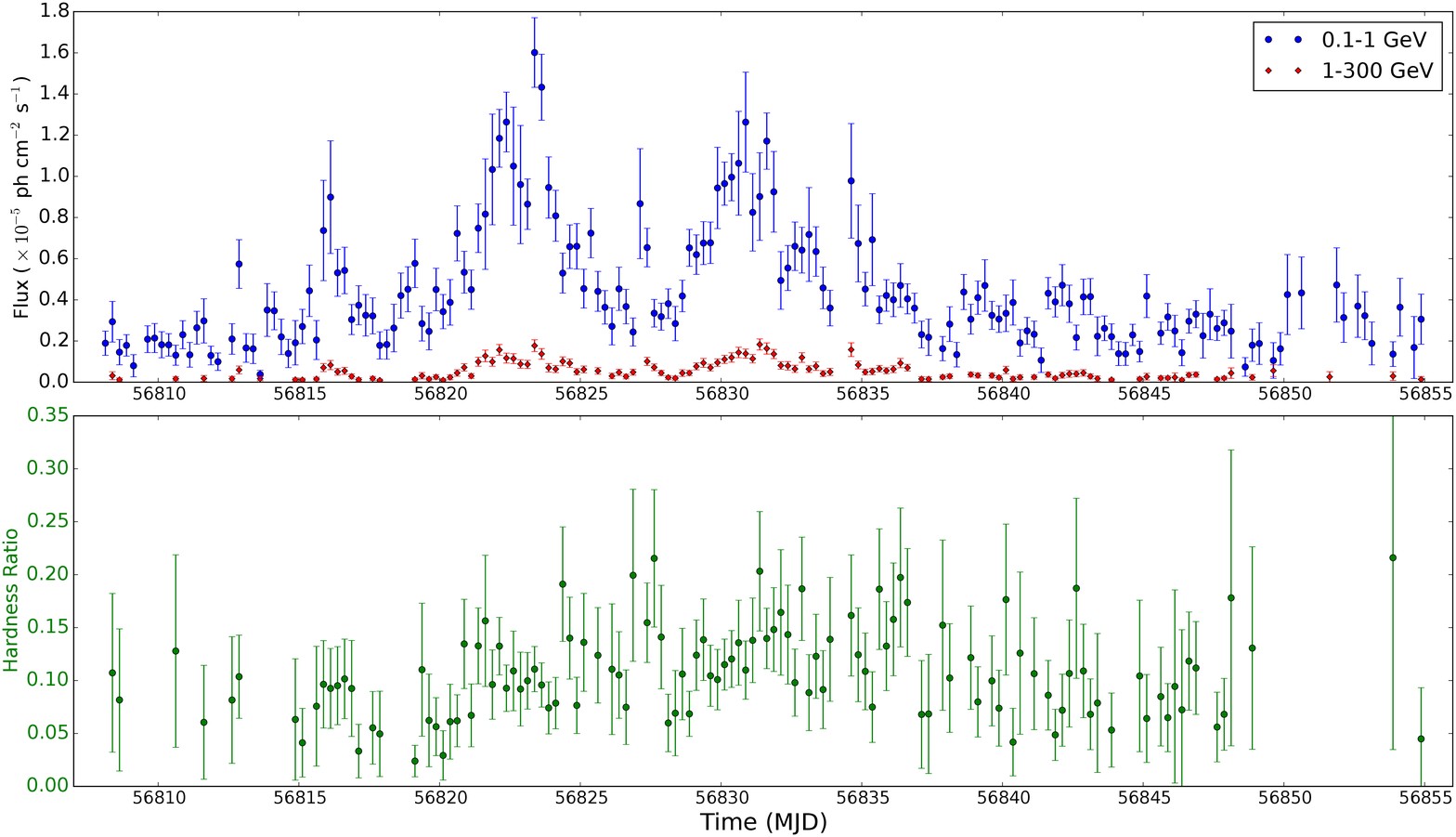}}
    \caption{Top: The light curve of 3C~454.3, in 6 hour time bins. The low energy flux, $0.1 \leq E_{\gamma}\leq1$~GeV is plotted using blue circles and the high energy flux, $1 \leq E_{\gamma} \leq 300$~GeV is plotted using red diamonds. Bottom: The corresponding hardness ratio of the $\gamma$-ray emission. This is the ratio of high energy flux to low energy flux. Only data points with $TS\geq10$ are shown.}\label{fig:hardness}
\end{figure*}

Fig.~\ref{fig:hardness} shows the $\gamma$-ray flux of 3C~454.3 in 6 hour time bins over the period of interest. The high energy, $1~\leq~E_{\gamma}~\leq~300$~GeV, and low energy, $0.1~\leq~E_{\gamma}~\leq~1$~GeV flux have been plotted separately, in order to highlight any energy-dependence that might exist in the rising and falling of the $\gamma$-ray flux. 6 hour bins were chosen to provide the balance between adequate statistics and being able to see the detail of the flare structure. The corresponding hardness ratio of 3C~454.3, the ratio of high energy flux to low energy flux, $F_{high}$/$F_{low}$, is also shown, and only bins with $TS\geq10$ are considered. No strong trend of hardness ratio with flux is identified, and the hardness ratio doesn't peak simultaneously with the flux. This may be evidence that the $\gamma$-ray flares are not solely due to an increase in flux at the high energy end of the spectrum.

As the energy-dependence of the flare cooling is an indicator of the energy-dependence of the emitting electron cooling, it is of interest to deduce whether or not any energy-dependence of the flare cooling exists, using Fig.~\ref{fig:hardness}. If the gradient of the hardness ratio is consistent with zero as the flares cool, the hardness ratio is remaining constant, meaning that the flux is not cooling differentially. This would indicate that the emission region is located within the BLR, with the IC scattering occurring in the KN regime \citep{ref:Dotson2012}. Conversely, energy-dependent cooling would manifest itself as a negative correlation of hardness ratio with time, according to our definition of hardness ratio and fig.~3 of \citet{ref:Dotson2012}. We perform a least-squares analysis between MJD 56823.2 and 56827 for the cooling of the first flare, and between MJD 56830.9 and 56834.5 for the second flare, to assess which of the two cases is applicable to our data. The results show that the hardness ratio is consistent with being constant as both flares cool, due to the large statistical uncertainties on the hardness ratio. It is also possible that substructures within the flares are masking any overall trend. Although it can be seen that there is variability in the hardness ratio between MJD 56808 and 56855, we would need less statistical uncertainty to come to a conclusion about the presence of energy-dependent cooling.

\section{Discussion}
\label{sec:discussion}
In order to draw conclusions on the location of the emission region, we need to combine the emission characteristics from all of the methods discussed in this paper.

The small amount of curvature and hard spectrum seen during the flares indicate an emission region that is not buried deep within the BLR. The optical depth calculations interpolated from \citet{ref:Pacciani2014} can be used to constrain the location of the emission region, since we observe significant emission of $E_{\gamma, emitted}\geq$~35~GeV and $E_{\gamma, emitted}\geq$~50~GeV photons. Assuming that the $\gamma$-rays are not oscillating to ALPs, optical depth arguments place the emission region at least $r\sim$1.3 $\times$ $R_{BLR}^{out}$ from the SMBH. \citet{ref:Pacciani2014} used MWL SED modelling to locate the emission region during a high energy activity period of 3C~454.3, in September 2013. They found that the emission region was located at $\sim$0.75~pc from the SMBH, which is significantly outside the BLR, upstream of the torus of 3C~454.3. They also found the $\gamma$-ray emission region to be located outside of the BLR for a number of other high energy FSRQs, and even discovered evidence of the emission originating downstream of the torus in two cases. MLW studies of the November 2010 flare from 3C~454.3, such as those by \citet{ref:Wehrle2012} and \mbox{\citet{ref:Vittorini2014}}, have also concluded that the favoured $\gamma$-ray emission model requires an emission location at parsec-scales from the SMBH. The $\gamma$-ray emission in the model suggested by \citet{ref:Vittorini2014} results from the scattering of photons reflected by a mirror cloud crossing the jet outflow. This was supported by the simultaneous variation in the optical continuum, highlighting the benefits of MWL studies.

We observe lower and more stable curvature during the flares compared to during the baseline emission either side, as seen in Figs \ref{fig:alphaBetaTime} and \ref{fig:betaFlux}. In addition to this, we observe the only significant $E_{\gamma}\geq$ 20~GeV emission during the flares. This could be interpreted as a different origin of the flare emission compared to that of the baseline emission. The KN suppression is mitigated as the emission region moves further downstream of the SMBH and BLR \citep{ref:Pacciani2014}, so the increased curvature during the baseline emission is consistent with both the increased KN suppression and the high energy attenuation from pair production in the BLR. It should, however, be highlighted that the magnitude of the KN suppression is also dependent on the emitting electron energy distribution \citep{ref:Tavecchio2008}. The increased curvature could be due to poor photon statistics either side of the flares, but the lack of high energy emission during the baseline emission strengthens the conclusion that the curvature is not due to statistics alone.

The change in emission characteristics during the flaring episode could indicate a multi-zonal emission model, where the baseline emission originates from inside the BLR and the flares from outside of the BLR. If we were to make more conservative conclusions based on the presence of the high energy emission, we could say that the flares must originate from the downstream half of the BLR or further. Previous studies such as those done by \citet{ref:Pacciani2010}, \citet{ref:Tavecchio2010}, \citet{ref:Bonnoli2011} and \citet{ref:Vercellone2011} have concluded that the $\gamma$-ray emission regions during the December 2009 and November 2010 flares of 3C~454.3 were located close to the SMBH. A long-term MWL campaign presented by \citet{ref:Vercellone2010} also concluded that the dominant emission mechanism of $\gamma$-rays from 3C~454.3 was the scattering of external photons around the BLR. A multi-zonal model for the flaring emission of 3C~454.3 may therefore also be applicable. Evidence for multiple emission regions, where emission originates in the MT and in the BLR simultaneously, was previously found by \citet{ref:Brown1510} for the FSRQ PKS~1510-089. This conclusion was primarily based on significant changes to the $\gamma$-ray spectral shape between flares separated by only a few days, and the seeming lack of correlation between the hardness ratio and the detection of high energy emission. It was concluded that one of the flares that \citet{ref:Brown1510} studied originated in the MT, based on the power law spectral shape and presence of high energy emission, similar to what we observe here for 3C~454.3.

Our investigation into the energy-dependence of the electron cooling did not reveal any significant differential cooling. This would indicate that the emission was originating from inside the BLR. Given the opposing evidence, and the fact that we identify variation in the hardness ratio across the flare period, finding no strong decreasing trend in the hardness ratio as the flares cool is most likely due to the high level of statistical uncertainty.

The short flux doubling timescales discussed in Section \ref{sec:taus} allow us to put an upper limit on the size of the emission region, $R\delta^{-1}<$~2.38 $\times$ $10^{-5}$~pc. Strikingly, there are no corresponding flux halving timescales that are less than 1.5~h. Assuming a leptonic model for the IC scattering, \citet{ref:Dotson2012} show in their fig.~2 that the cooling of electrons in the BLR is much faster than in the MT, for a given energy. This is because the IC scattering occurs under the Thomson regime in the MT rather than the KN regime, due to the lower external photon field energies. The comparatively slow cooling that we see during the 2014 flares of 3C~454.3 is therefore also in support of an emission region that is not inside the BLR.

The size of the BLR, $R_{BLR}^{out}$, in 3C~454.3 is $\sim$0.2~pc \citep{ref:Bonnoli2011}. Therefore, the cross-sectional diameter of the jet at $r=1.3 \times R_{BLR}^{out}$ ($r\sim$0.26~pc), is $\sim$0.05~pc if we assume a constant opening angle of $\sim$0.1 rad \citep{ref:GeneralBlazars}. Comparing this to the calculated size of the emission region, $R\delta^{-1}<$ 2.38 $\times$ $10^{-5}$~pc, the jet at this point is $\sim$2 orders of magnitude too large for the emission region to be covering the cross-section of the jet. This calculation of jet diameter does however assume that the geometry of the jet is constant and cone-like. Studies such as \citet{ref:Marscher2006, ref:Villata2007, ref:Vercellone2010} and \citet{ref:Mizuno2015} have suggested that this geometry is not the case for all relativistic jets of AGN, and that the jets may in fact bend or re-collimate in some cases. Structural observations of the jet are difficult when they are directed so closely towards our line of sight, so it may be the case that the geometry of the jet of 3C~454.3 is also not constant. If the jet of 3C~454.3 does re-collimate, the diameter of the jet at the location of the emission region may be smaller than for our assumed geometry.

We suggest that the $\gamma$-ray emission region for the June 2014 flares can be well described as a blob-in-jet outside of the BLR. This is consistent with the short $\gamma$-ray flux doubling timescales, the relatively long flux cooling timescales, the presence of significant high energy emission and the low amount of curvature present in the $\gamma$-ray spectrum during these flares. Characterising the emission region as covering the entire cross-section of the jet would also be consistent with observations, if we believe that the jet can re-collimate to the size of the emission region downstream of the BLR.

\section{Conclusions}
\label{sec:conc}
In this study, we use data collected by the \textit{Fermi}-LAT to examine the 0.1 $ \leq E_{\gamma} \leq$ 300~GeV $\gamma$-ray emission characteristics of 3C~454.3. We isolate the period between MJD 56799 and 56855, during which the FSRQ underwent a bright flaring episode spanning $\sim$25 days. The $\gamma$-ray flux doubling timescales were calculated during the period of interest by binning the data into good time intervals, whilst maintaining a $TS\geq10$ selection criterion. Four intrinsic doubling timescales $\tau_{int}$ < 1~h were found, with a fastest doubling timescale of $\tau_{int}$ = 0.68 $\pm$ 0.01~h. This allows us to calculate an upper limit on the size of the emission region of $R\delta^{-1}<2.38 \times 10^{-5}$~pc.

The $\gamma$-ray spectral shape, evidence for high energy emission and the energy-dependence of the electron cooling were investigated in order to constrain the distance of the emission region along the jet. We find the spectral curvature during the flares to be low and steady in comparison with the baseline $\gamma$-ray emission. We also observe significant $E_{\gamma, emitted}\geq$~35~GeV and $E_{\gamma, emitted}\geq$~50~GeV emission from 3C~454.3 over the flaring period. Optical depth calculations therefore allow us to constrain the position of the emission region to be outside of the BLR, at $r\geq1.3$~$\times$~$R_{BLR}^{out}$ from the SMBH.

The spectral changes when the flares erupt and the lack of $E_{\gamma}\geq$ 20~GeV emission either side of the flares lead us to believe that the baseline emission may be originating from a different part of the jet than the flares. The curvature before and after the flares is larger than during the flares, but better photon statistics during the baseline emission are required to investigate this idea further. We identify variation in the hardness ratio of the $\gamma$-ray flux, but we find no evidence of differential flux cooling times. We are, however, too limited by statistical uncertainty to make firm conclusions. 

We conclude that the flaring emission region is located outside the BLR. Due to the compact size of the emission region, the emission region is either a blob-in-jet or distributed across the cross-section of the jet, depending on the jet geometry. These conclusions differ from the traditional view of $\gamma$-ray emission, both from 3C~454.3 and AGN more generally. They are, however, in support of more recent studies that have reported non-BLR emission from FSRQs \citep{ref:Brown1510, ref:Dotson2015} and 3C~454.3 in particular \citep{ref:Wehrle2012, ref:Vittorini2014}. We suggest that 3C~454.3 could be another example of an FSRQ that emits $\gamma$-rays from multiple emission regions. Future work may allow us to gain better photon statistics during the baseline emission, and would enable us to make conclusions on the location of the baseline emission in order to further explore the possibility of a multi-zonal model.

\section*{Acknowledgements}

We thank the referee for their comments and suggestions that have improved the quality and clarity of this paper. We also thank G. Bonnoli and collaborators for kindly sharing their data from previous states of 3C~454.3. This work has made use of public \textit{Fermi} data obtained from the High Energy Astrophysics Science Archive Research Center (HEASARC), provided by NASA Goddard Space Flight Center. AMB acknowledges the financial support of the University of Durham.




\bsp	
\label{lastpage}
\end{document}